\documentclass [12pt]{article}

\begin{document}

\title{On the Problems of Hazardous Matter and Radiation at Faster than Light Speeds in the Warp Drive Space-time}

\author{C. B. Hart \and R. Held\footnote{ronaldheld@aol.com} \and P. K. Hoiland \and S. Jenks \and F. Loup\footnote{loupwarp@yahoo.com} \and D. Martins \and J. Nyman \and J. P. Pertierra\footnote{pertierr@purdue.edu} \and  P. A. Santos \and M. A. Shore \and R. Sims \and M. Stabno \and T. O. M. Teage}

\maketitle

\paragraph{Abstract: }
The problems of hazardous radiation and collisions with matter on a warp driven ship pose considerable obstacles to this possibility for interstellar travel.  A solution to these problems lies in the Broeck metric.  It will be demonstrated that both threats to the ship will be greatly reduced.  It will also be shown that the horizon problem no longer exists.

\section{Introduction}
A warp driven vehicle travelling at a speed faster than light may collide with objects in front of the ship, which would be hazardous to the ship and its crew.  Additionally, another problem exists in that photons arriving at the front of the ship are blueshifted to very high energies by the Pfenning warped region.  This high energy radiation can be lethal to the ship's crew and damaging to the ship itself.  It is desirable to find a solution that protects the ship from both of these threats while mantaining the warp drive characteristics required for FTL travel.  It is also desirable that this solution permits signals to be sent from the ship to the outside world.

\section{Broeck Warp Drive metric and Broeck warped regions}
The warp drive metric in \cite{Broeck} was used to create a new metric.  This variant of the Broeck warp drive metric differs from the previous metric of \cite{Paper1} in that it contains two warped regions. One is the usual Pfenning warped region representing the behavior of the Top Hat function $f(rs)$ going from $1$ to $0$, and the other is the Broeck warped region, which will slow down incoming photons in the neighborhood of the ship.  For the remote frame, the new metric is:  

\begin{equation}
\label{eq1}
ds^2= 1-B^2[dx-v_sf(r_s)dt]^2
\end{equation}

In the ship frame, the metric can be written as:

\begin{equation}
\label{eq2}
ds^2 = 1-B^2[dx'+v_sg(r_s)dt]^2.
\end{equation}

The symbols have the following definitions:

\begin{equation}
\label{eq3}
f(r_s)=\frac{tanh[\delta(r_s+R)]-tanh[\delta(r_s-R)]}{2tanh[\delta(R)]}
\end{equation}

$f(r_s)$ is the Alcubierre Top Hat related to Pfenning Warped region $R-\delta$ to $R+\delta$.

\begin{equation}
\label{eq4}
B=[\frac{1+tanh[\delta(r_s-D)]^2}{2}]^{-P}
\end{equation}

Where $D$ is the radius of the Broeck warp region, $R$ is the radius of the warp bubble. 

\begin{equation}
\label{eq5}
r_s=\sqrt{[(x-x_s)^2 + y^2 + z^2]}
\end{equation}

\begin{equation}
\label{eq6}
v_s=\frac{dx_s}{dt}
\end{equation}

\begin{equation}
\label{eq7}
x'=x-x_s
\end{equation}

\begin{equation}
\label{eq8}
g(r_s)=1-f(r_s)
\end{equation}

Two alternative(non-hyperbolic) continuous behaviour expressions for $f(r_s)$ and $B$
are presented here:

\begin{equation}
\label{eq9}
f(r_s)=\frac{1}{[(\frac{r_s}{R})^(2(R+1))+1]}
\end{equation}

\begin{equation}
\label{eq10}
B(r_s)=\frac{2}{(\delta[(r_s-D)^(2(D+1))+1])}+1
\end{equation}

The general behavior of $B$ is $1$ at the ship and far from it, while large in the vicinity of $D$.  The exponent $P$ is a free parameter.  A more convenient form of the metric is:

\begin{equation}
\label{eq11}
ds^2=[1-(Bv_sg(r_s))^2]dt^2-2v_sg(r_s)(B^2)dx'dt-(B^2)dx'^2
\end{equation}

\section{Behavior of Photons in Ship Frame}

Solving equation \ref{eq11} for photons($ds^2=0$), produces two solutions:

\begin{equation}
\label{eq12}
v_1=-v_sg(r_s)+\frac{1}{B} 
\end{equation}

\begin{equation}
\label{eq13}
v_2=-v_sg(r_s)-\frac{1}{B}
\end{equation}
 
For the rest of this paper, $v_1$ will represent photons being emitted in the direction of motion, and $v_2$ for photons emitted in the opposite direction, as seen from the ship. Finally, $v_3$ is the speed of a photon coming towards the ship from the front as:

\begin{equation}
\label{eq14}
v_3=-v_sg(r_s)-\frac{1}{B}
\end{equation}

The qualitative analysis of the photon's speed can be broken up into five regions:  near the ship, the Broeck region, between the Broeck and Pfenning regions, the Pfenning region, and finally outside of the Pfenning region. At the location of the ship $g(r_s)=0$, and $B=1$, so $v_1=1$ and $v_2=-1$. At a distance $D$ from the ship, in the Broeck warped region, $g(r_s)=0$, and $B$ is large. For a large enough $B$, both $v_1$ and $v_2$ can be made proportionally small due to $v_1=+1/B$ and $v_2=-1/B$.
In between the Broeck and Pfenning warped regions, the space is effectively flat, so $g(r_s)=0$, and $B=1$, therefore $v_1=1$ and $v_2=-1$.
In the Pfenning Warped region, $g(r_s)$ is between $0$ and $1$. Here, $v_2$ is still are negative, but $v_1$ changes sign at $g(r_s)=1/v_s$.
Finally, outside of the Pfenning warped region, $g(r_s)=1$ and $B=1$. This yields:

\begin{equation}
\label{eq15}
v_1=-v_s+1 
\end{equation}

($v_1$ changes sign)

\begin{equation}
\label{eq16}
v_2=-v_s-1 
\end{equation}

Note again that the photons entering the Pfenning region will be accelerated. The Broeck warped region is designed to slow them down.  Observe that the velocity of the photon going backward never changed sign. The rear of the warp bubble and the external regions are causally connected to the ship, although these regions cannot send signals to the ship. 
It is a different situation for the photons going forward. For speeds greater than $1$, the photon will have a negative speed of propagation moving forward.  A Horizon occurs when the photon changes the direction of the speed. 
This important topic will be discussed later.

\section{Incoming Photons in the Broeck Warped Region}

As was stated in previous Section, a photon coming towards the ship from the front, has speed $v_3$ defined as:

\begin{equation}
\label{eq17}
v_3=-v_sg(r_s)-\frac{1}{B}
\end{equation}

$B$ will be large at a distance of $D$ from the ship.  At $D$, $B$ will have its maximum value, while being $1$ in the ship, and far from it. If the warp bubble moves with speed $v_s=100$, photons will come to the ship with a relative speed $v_3=-101$. When they enter the region where $B$ is large, the photons will decelerate to reach a minimum value at $D$ due to $v_3 = -1/B$ when $g(r_s)=0$, inside of the bubble.  By placing $D$ relatively close to the ship, photons or incoming particles can be slowed down in the vicinity of the ship, reducing the danger of collisions.

\section{Behavior of photons emitted by the Ship in the Remote frame}
 
In the remote frame and in front of the ship, the metric is:

\begin{equation}
\label{eq18}
ds^2=1-B^2[dx-v_sf(r_s)*dt]^2 
\end{equation}

or after expanding

\begin{equation}
\label{eq19}
ds^2=[1-(Bv_sf(r_s))^2]dt^2+2v_sf(r_s)(B^2)dxdt-(B^2)dx^2.
\end{equation}

Solving the equation for the speed of the photons, for $ds^2=0$, produces two solutions:

\begin{equation}
\label{eq20}
v_1=v_sf(r_s)-\frac{1}{B}
\end{equation}

\begin{equation}
\label{eq21}
v_2=v_sf(r_s)+\frac{1}{B}
\end{equation}
 
In this section, $v_2$ will represent photons being emitted in the direction of motion, and $v_1$ is the opposite of $v_2$. These photons are emitted by the ship, but seen by the remote frame in front of it(this is important).
As was done for the ship's frame, the qualitative analysis of the photon's speed can be broken up into five regions. At the location of the ship $f(r_s)=1$, and $B=1$, so $v_2=1+v_s$ and $v_1=v_s-1$. In the Broeck warped regions, $f(r_s)=1$, and $B$ is large so the square root of $B$ will be large. For large enough $B$, both $v_1$ and $v_2$ are approximately equal to $v_s$. In between the Broeck and Pfenning warped regions, $f(r_s)=1$, and $B=1$, so $v_2=1+v_s$ and $v_1=-1+v_s$. In the Pfenning Warped region, $f(r_s)$ is between $0$ and $1$, $B$ is $1$. Here, $v_2$ and $v_1$ still are positive and negative, respectively.  Finally, outside of the Pfenning warped region, $f(r_s)=0$, and $B=1$. This yields:

\begin{equation}
\label{eq22}
v_2=1
\end{equation}

\begin{equation}
\label{eq23}
v_1=-1
\end{equation}
 
Notice that the photons going forward will leave the warped space and be detected by an observer far in front of the ship. This indicates that information from the ship can be sent to the front of the warp bubble. This contrasts with the observer on the ship, who loses contact with the photons in a part of the Pfenning warped region.
The photons emitted in the forward direction, when seen by the ship, will change sign of speed in the horizon and the ship will loose contact with the photons. The remote frame will see the photons emitted by ship, crossing the horizon and emerging outside.
This behavior is similar to the event horizons of black holes, in which a remote observer never sees the photons crossing the event horizons but an observer inside the hole will see the photons go into the singularity.
Previous papers stating that photons sent by the ship towards the direction of travel never reach the outermost layers of the bubble must be reexamined.  The ship loses contact with the photons in the horizon, but the photons carrying the information will be seen by an outside observer in front of the ship emerging from the warp bubble and reaching the external space-time. 

\section{The behavior of matter coming towards the ship}

The exact description of what happens to matter entering the space-time is quite complex. Since a time-like interval is appropriate to matter, the equations in section 2 would have the = replaced by < for equations with $v_3$. The descriptions are still qualitatively the same. 
Here, we offer a qualitative explanation, with more details to follow in a future work.  Pieces of matter too small to be disrupted by the tidal forces will be slowed down in the Broeck warped region, potentially to impact the ship at relatively slow speeds. For larger pieces of matter, they will become tidally disrupted in the Broeck regions. The trajectories of the pieces are complicated, but the pieces are not expected to impact the ship at high speeds.

In the vicinity of the ship (Pfenning Region) (ship frame)

\begin{equation}
\label{24}
v_3<-v_sg(r_s)-1/B 
\end{equation}

but in the neighborhoods of the ship $g(r_s)=0$ then

\begin{equation}
\label{25}
v_3<-1/B  
\end{equation}

For a large $B$, $v_3$ will be low.

Refine $B=B_1B_2$ where:

\begin{equation}
\label{eq26}
B_1=[\frac{1+tanh[\delta(r_s-D)]^2}{2}]^{-P}
\end{equation}

\begin{equation}
\label{eq27}
B_2=[\frac{1+tanh[\delta(r_s-S)]^2}{2}]^{-P}
\end{equation}

$S$ is a distance from the ship where $S > R$.

$B_1$ is $1$ at the ship and far from it, while is large at a distance $D$ from the ship. $B_2$ is $1$ at the ship and far from it while large at a distance $S$ from the ship. This produces two regions where the matter is slowed down, one at $D$ and the other at $S$.
$S$ outside Pfenning region but involving Pfenning Region as a protection layer will slow down incoming matter that otherwise would enter Pfenning Region at high speeds and disrupt possible the Warp Field.
For a rigid body, the gravitational gradients in Broeck regions will tidally disrupt hazardous objects in the ship's neighborhood. This is a property of Broeck space-time, any natural object will be disrupted and deflected (if not on a head-on collision). 

qualitative description seen from a remote frame
for a incoming particle or photon approaching the ship from the front.

\begin{equation}
\label{eq28}
v_3 =  v_sf(r_s) - \frac{1}{B}
\end{equation}

Outside the region $S$, $B$ is $1$ and the speed of the
incoming photon seen by a remote frame is $v_3=-1$
outside Pfenning region at a distance $S$ from the ship,
$S>R$, $f(r_s)=0$, and $v_3 = -1/B$ at a
distance $S$.  Therefore meteors and debris will be slowed
down before entering Pfenning Warped region.

A photon sent forward by the ship but seen by a remote
frame outside Pfenning region in the vicinities of $S$

$v_2=1/B$ the photon seen by the remote frame never
changes sign so the photon sent by ship but seen by
the remote frame will cross Pfenning region and the
outermost Broeck region destinated to protect Pfenning
region from impacts and arrive at normal space.

\section{Energy needed to sustain a warp field}
Recently Krasnikov \cite{Krasnikov} demonstrated that a Broeck warp drive can be obtained with approximately -10 kg of exotic matter if we use classical scalar fields instead of quantum fields. This was also noted in the papers by Ford-Roman  \cite{Ford-Roman}, Visser-Barcelo \cite{Visser-Barcelo} ,and Lobo-Crawford \cite{Lobo-Crawford}.
In the paper of Ford-Roman 2000, they found at equation 15 an energy of $-3*10^17 gm cm^2/s^2$, this amount can be generated using a scalar massless non-minimally coupled scalar field. Krasnikov reduced the amount of energy needed for a macroscopic sized radius of $-10kg$. If his assumptions are correct, the creation of warp fields will not require astronomical amount of negative energy.

\section{Conclusion}
In the previous sections, we have demonstrated that the Broeck space-time properties advance the knowledge need to creating a functional warp drive. Assuming that space-time can be manipulated to create warp driven spacecraft, there are other applications possible,
One such application is the deflection of matter in the form of meteors, away from ships, space stations, or sections of planets. The space-time properties that are intrinsic to the warp metric can be used to deflect matter. If this could be made practical, meteors could be deflected away from planets via sub-light ships. Perhaps one could create a static warp field, which would perform the same function.  In either case, this is another avenue of research to be investigated.
Using a variant of the original Broeck metric, we have demonstrated that the incoming radiation will not pose a major problem to the ship, as well as the impinging matter. In addition, the warped regions will retain the capability to receive signals from the ship. 
The authors believe that the warp drive space-times cannot be ruled out by the limitations of energy, blue shifts or horizons. Furthermore, the ability to generate warp drive space-times permits other applications other than propulsion.

\end{document}